\documentclass[twocolumn,showpacs,amsmath,amssymb,superscriptaddress]{revtex4}
\usepackage{graphicx,url,hyperref,dcolumn,bm,lineno}
\hypersetup{colorlinks,citecolor=blue,linkcolor=red,urlcolor=blue}
\usepackage{graphicx}

\begin{document}
%--------------------------------------------------------------------
\title{Local statistics of immiscible and incompressible two-phase flow in porous media}
%--------------------------------------------------------------------
\author{Hursanay Fyhn}
\email{hursanay.fyhn@ntnu.no}
\affiliation{PoreLab, Department of Physics, Norwegian University of Science and Technology, NTNU, N-7491 Trondheim, Norway}

\author{Santanu Sinha}
\email{santanu.sinha@ntnu.no}
\affiliation{PoreLab, Department of Physics, University of Oslo, N-0316 Oslo, Norway}

\author{Alex Hansen}
\email{alex.hansen@ntnu.no}
\affiliation{PoreLab, Department of Physics, Norwegian University of Science and Technology, NTNU, N-7491 Trondheim, Norway}

%--------------------------------------------------------------------
\date{\today {}}
%--------------------------------------------------------------------
\begin{abstract}
We consider immiscible and incompressible two-phase flow in porous media under steady-state conditions using a dynamic pore network model. We focus on the fluctuations in a Representative Elementary Area (REA), with the aim to demonstrate that the statistical distributions of the volumetric flow rate and the saturation within the REA become independent of the size of the entire model when the model is large enough. This independence is a necessary condition for developing a local statistical theory for the flow, which in turn opens for the possibility to formulate a description at scales large enough for the typical pore size to be negligible using differential equations. 
\end{abstract}
%--------------------------------------------------------------------
\maketitle
%--------------------------------------------------------------------
% \tableofcontents
%--------------------------------------------------------------------
\section{Introduction}
\label{intro}

When two or more immiscible fluids compete for space while flowing in a porous medium, we are dealing with multiphase flow \cite{b88,s11,b17,ffh22}. Finding a proper description of multiphase flow at the Darcy scale, which may be orders of magnitude larger than the pore scale, is a central problem in porous media research.  On the Darcy scale, the only practical approach to the multiphase flow problem is to replace the original porous medium with a continuous medium and then describe the flow through a set of differential equations relating the fluid velocities to the driving forces, e.g. pressure gradients, saturation gradients and gravity. The approach dominating any practical applications of immiscible two-phase flow that today requires calculations is based on relative permeability theory \cite{wb36}. This is a purely phenomenological theory essentially stating that the two immiscible fluids get into each other way and therefore reduce the effective permeability each fluid experiences.  Add a capillary pressure function to take into account the capillary forces between the two fluids, and the theory is complete \cite{l40}.  This phenomenological approach has the flaw that it provides no path to implement into it our increasing understanding of the interactions and flow of the fluids at both the pore scale and the molecular scale.  

Solving the {\it scale-up problem\/} in immiscible two-phase flow in porous media consists of expressing the flow at the pore scale in terms of the flow at the molecular scale and then expressing the flow at the Darcy scale in terms of the flow at the pore scale. The favored approach to the scale-up problem is that of homogenization. That is, start with a description of the problem on small scale using variables appropriate for that scale. Then average these variables over the large scale, followed by closure assumptions.

One example of the homogenization approach of scaling up immiscible two-phase flow in porous media starts from  mechanical principles such as momentum conservation to arrive at an effective description of the flow through homogenization \cite{w86,as86,a87,alb89,pb18,lv22}. Thermodynamically Constrained Averaging Theory (TCAT) \cite{hg79,hg90,hg93,hg93b,nbh11,gm14} is a very different approach to the scale-up problem. It is based on volume averages of thermodynamic quantities defined at the sub-pore and pore scale, together with closure relations at the homogeneous scale as formulated by Whittaker \cite{w86}. Another homogenization approach is that of Kjelstrup et al.\ \cite{kbhhg19,kbhhg19b,bk22}, who use Euler scaling to work out the averages of intensive variables such as pressure. This approach manages to keep the number of variables down in contrast to other approaches. We also point to the homogenization approach based on expressing the central thermodynamic potentials in terms of geometric variables that characterize the porous medium, the fluid interfaces and the contact lines and the Minkowski functionals combined with powerful theorems from differential geometry \cite{cacbsbgm18,acbrlasb19,cba19,cfbablr22}.

These homogenization approaches succeed in taking the description of the flow from the sub-pore scale to scales just above the pore scale.  They do not, however, take into account the fluid structures that appear at even larger scales. These structures result from the way the fluids arrange themselves within the porous medium, i.e., their cluster structure.  They profoundly affect the flow on the intermediate scales below the Darcy scale --- and this must be reflected in the flow at the Darcy scale. Energetically, these structures are not dominating, and therefore easy to discard in the different homogenization approaches. 
However, any scale-up attempt taking the problem from the pore scale to the Darcy scale needs to take these structures into account.  As the structures appear over many length scales, a different approach from those based on homogenization techniques is needed. 

Looking back in history, there is an upscaling technique that is capable of dealing with structures and correlations that stretch across scales: statistical mechanics \cite{r76}. The early developers of thermodynamics constructed their approach in order to understand heat and its relation to work in parallel to the development of the steam engine. It is based on conservation laws and symmetries, especially dilation symmetry.  It treats the medium as a continuum and provides the necessary differential equations. Statistical mechanics was developed to understand how the motion of atoms and molecules leads to the thermodynamic relations, i.e., it provides the scaling up from the molecular scale to the continuum scale, thus circumventing the necessity to solve the equations of motion for every molecule.

One may therefore get the impression that thermodynamics and statistical mechanics are inextricably linked to atomic and molecular systems. This is, however, not correct. 
Jaynes \cite{j57} developed a generalized statistical mechanics in the fifties based on the statistical approach to information developed by Shannon a few years earlier \cite{s48}. 
This approach, in turn, originates in the {\it principle of sufficient reason\/} formulated by Laplace \cite{l51}: 
If we know nothing about a process with two outcomes, the optimal choice of probabilities for the two outcomes is 50~\% for each. 
Shannon constructed a {\it function of ignorance\/} measuring quantitatively what we do {\it not\/} know about a given process having a number of different outcomes. One of his criteria for this function, called the Shannon entropy, was that it would have its maximum value when the probabilities for all outcomes would be equal, which is a generalization of the Laplace principle of sufficient reason. 
Jaynes took this approach further by adding the criterion that the Shannon entropy is maximum given what is known about the process. This leads to a set of equations that determine the probabilities for the different outcomes. This is Jaynes' generalization of statistical mechanics.  

An important caveat in applying the Jaynes maximum entropy approach is that it does not work for driven systems \cite{cd19}. Immiscible two-phase flow in porous media does represent a driven system where there is production of entropy due to viscous dissipation and irreversible motion of fluid interfaces and contact lines. Nevertheless, in a recent paper, Hansen et al. \cite{hfss22} developed a statistical mechanics for immiscible and incompressible two-phase flow in porous media based on the Jaynes principle of maximum entropy, leading to a formalism resembling thermodynamics that describes the flow at the continuum level. The trick to make it work was not to consider the molecular entropy which is being produced when the fluids move, but rather the entropy associated with the flow patterns of the fluids. This entropy is {\it not\/} being produced under steady-state flow conditions. Furthermore, it is this entropy that properly describes the fluid structures on scales above the pore scale, whereas the molecular entropy associated with dissipation dominate at scales up to the pore scale. 

The Jaynes approach solves the scale-up problem in the same way as it was solved through ordinary statistical mechanics for atomistic systems. It is the aim of the present paper to investigate numerically a necessary criterion which was only assumed to be true in \cite{hfss22} for the Jaynes approach to be applicable to immiscible two-phase flow in porous media: can we partition the porous medium into a ``system" in contact with a ``reservoir" as in ordinary thermodynamics? The term ``reservoir" has very different meanings in thermodynamics and in porous media research.  In this work, the term is used in a thermodynamical sense, which is that a reservoir is a system large enough so that the variables describing it do not change when brought into contact with a system small enough for its variables to be affected. The way we answer the question just posed is this: Based on a numerical model, we record the statistics of key parameters in the system for different sizes of the reservoir, finding that the statistics is independent of the reservoir size when it is large enough. 

We note that there have been earlier attempts at capturing the evolution of retention in unsaturated porous media subject to quasi-static changes in imposed pressure. Xu and Louge \cite{xl15} formulate drainage or imbibition through porous media using an Ising model that predicts the retention curve of saturation vs capillary pressure.
This is a very different approach with different aims from that of Hansen et al.\ \cite{hfss22} who focus on steady-state flow.  

We will in the following relate the concept of a ``system" to that of a {\it Representative Elementary Area\/} (REA) \cite{bb12}. 
At each point in the pore space of the porous medium, we may place an area that is orthogonal to the streamline passing through it.  
The area qualifies as an REA if it is large enough for the variables describing the properties of the medium itself and the fluids passing through it to have well-defined averages.
To obtain meaningful averages, the length scale of REA must be larger than the microscopic characteristic length of the porous medium to avoid rapid small-scale fluctuations, and must also be smaller than the characteristic length of the large-scale inhomogeneities \cite{hg79,rr14}.

The statistical mechanics developed by Hansen et al.\ for immiscible and incompressible two-phase flow in porous media \cite{hfss22}, leading to a thermodynamics-like formalism for the macroscopic variables describing the flow \cite{hsbkgv18,rsh20,rpsh22,ph22}, is reviewed in Section~\ref{Jaynes}.  We go into some detail here in order to place the present work in a proper context.

The dynamic pore network model \cite{ambh98,sgvh19} used in this work is introduced in Section~\ref{dpnm}.  The model is implemented as a two-dimensional lattice where the REA is defined as a one-dimensional sub-lattice placed orthogonally to the average flow direction.

The aim of this paper is to demonstrate that Equation~\eqref{eq-4} is valid for our dynamic pore network model.  This equation states that the statistics of the variables characterizing the REA do not depend on the statistics of the reservoir apart from local interactions. We report on our findings in Section~\ref{system}.  
We first investigate how the statistics of the variables we focus on, the wetting saturation and the Darcy velocity, vary with the size of the sub-lattice we consider, see Subsection~\ref{sysVarStat}. This allows us to determine when the sub-lattice is large enough to act as an REA. 
We then proceed to study the dependence of the variable fluctuations on the size of the REA in Subsection~\ref{stdev}. Surprisingly, whereas the fluctuations of the wetting saturation, scale as the inverse of the square root of the size of the REA, the average Darcy flow velocity fluctuations scale as the inverse of the size of the REA to the power 0.83.    
Lastly, in Subsection~\ref{independence} we test whether the statistics measured in the REA are independent of the size of the reservoir.  We do indeed find that this is, thus verifying the validity of Equation~\eqref{eq-4} for our dynamical pore network model.  

A pertinent question is, what would happen if the verification of Equation~\eqref{eq-4} would have failed?  It would invalidate the statistical mechanics of Reference \cite{hfss22}, but it would also have a negative impact on any attempt at constructing a local theory for immiscible two-phase flow at the Darcy scale in that all quantities are local. Rather than having the theory represented in the form of differential equations, they would contain integrals over space.  We summarize and discuss this in Section~\ref{conclusion} in addition to the other results.

%--------------------------------------------------------------------
\section{Statistical Mechanics}
\label{Jaynes}

\begin{figure}[ht]
\centering
\includegraphics[width=0.55\linewidth]{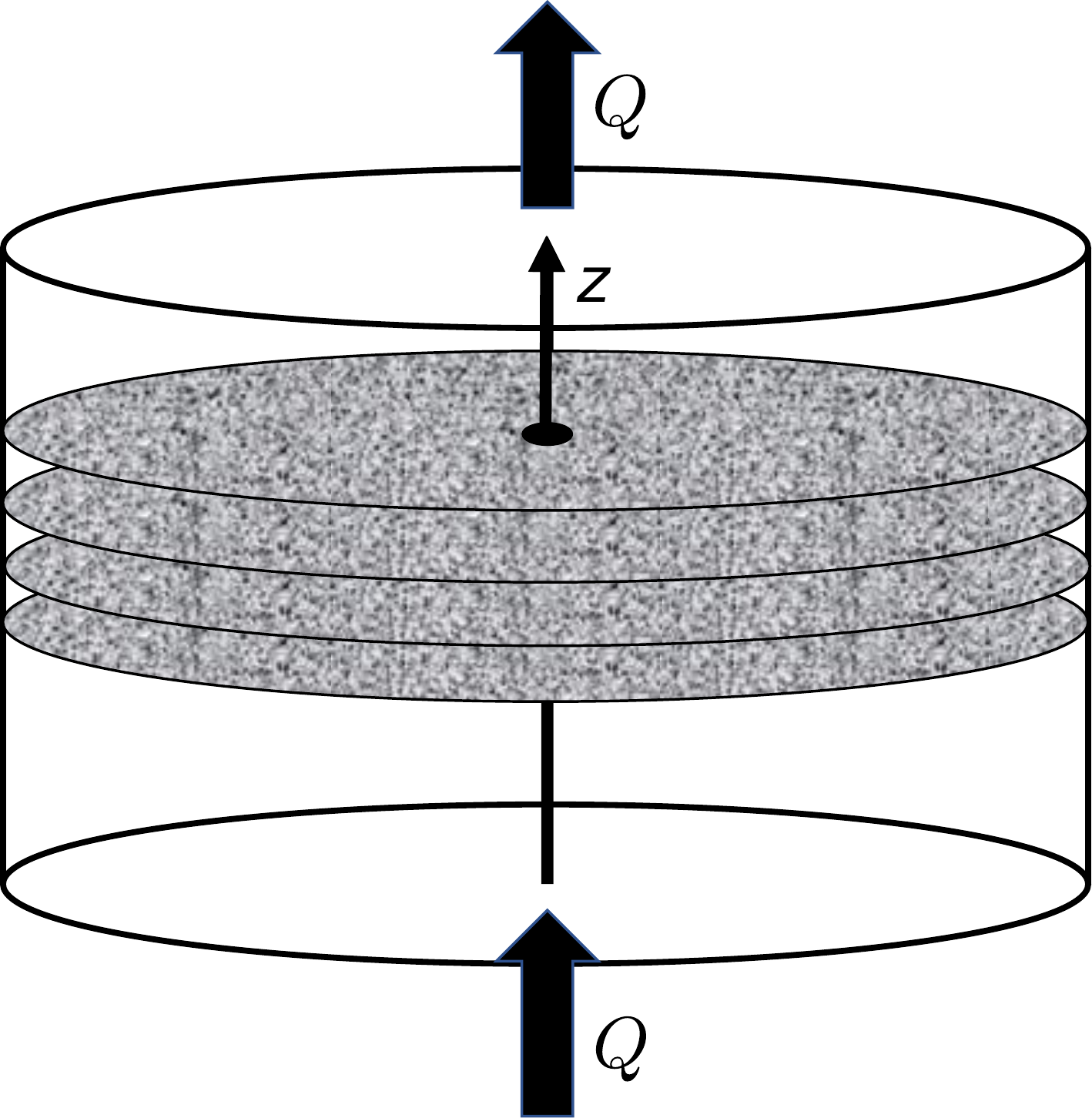}
\caption{A porous medium in the shape of a cylinder. 
  There is a flow of volumetric flow rate $Q$ passing through it.
  Four planes cutting through the cylinder orthogonally to the average flow direction, i.e., the $z$-axis, are shown. The volumetric flow rate $Q$ is the same through each plane.  However, the volumetric flow rate of each fluid, $Q_w$ and $Q_n$ vary from plane to plane.}
\label{fig1}
\end{figure}

We review in the following the statistical mechanics approach to immiscible and incompressible two-phase flow in porous media of Hansen et al.\ \cite{hfss22}.
Envision a homogeneous cylindrical block of porous medium as shown in Figure~\ref{fig1}, with a volumetric flow rate $Q$ flowing through it.
This flow consists of two immiscible and incompressible fluids which are well mixed before entering the porous medium. Keeping the flow entering into the porous media constant creates a steady-state flow within the porous medium. 
By steady-state flow we mean that the macroscopic variables describing the flow remain constant or fluctuate around well-defined averages. It is important to note that this does not imply that the pore scale interfaces between the fluids remain static. Rather, at the scale of the fluid clusters, there may be strong activity where clusters form and break up. Steady-state flow is a concept that is defined at the macroscopic Darcy level, not at the pore level. We may split $Q$ into the volumetric flow rate of the more wetting fluid, $Q_w$, and the volumetric flow rate of the less wetting fluid, $Q_n$, so that
\begin{equation}
\label{eq-1}
Q=Q_w+Q_n\;.
\end{equation}

%--------------------------------------------------------------------
%--------------------------------------------------------------------

The flow is dissipative and hence molecular entropy is produced.   
There is viscous dissipation and the motion of fluid interfaces and contact lines contains a dissipative element \cite{m70}.  This means that there is a production of entropy as hydrodynamic motion is converted into thermal motion.  The Jaynes maximum entropy principle should therefore not be applicable \cite{cd19}.  We now explain how we get around this hurdle.

There are three scales that stand out in porous media: the molecular scale, the pore scale and the Darcy scale.  At the sub-pore scale, the dissipation dominates the flow and methods from non-equilibrium thermodynamics are appropriate \cite{kbhhg19,kbhhg19b,bk22}. However, on scales above the 
pore scale, it is the fluid clusters and how they move that dominate. One may associate an entropy with these fluid structures.  

In order to construct this {\it flow entropy,\/} we imagine a cylindrical porous plug as shown in Figure~\ref{fig1}.  There is immiscible two-phase flow in the direction of the cylinder axis. We now focus on a set of imaginary planes that cut through the porous plug orthogonal to the cylinder axis as shown in Figure~\ref{fig1}. Imagine each plane is divided into voxels with sufficient resolution. Each voxel is associated with a number of variables describing the flow through it.  To be concrete, suppose we model the porous medium using the Lattice Boltzmann method (LBM) \cite{rbt19}. The voxels would then be the nodes of the lattice along the plane used in LBM and the variables would be the LBM variables associated with these nodes. 
The configuration in the plane, $X$, would be the values the voxel variables have at that particular instance in each voxel.
Measuring over many configurations we may define a {\it configurational probability density\/} $P(X)$.  This, in turn, defines our entropy,
\begin{equation}
\label{eq-2}
\Sigma=-\int\ dX\ P(X)\ \ln P(X)\;,
\end{equation}             
where the integral is over all {\it physically feasible configurations\/} in the plane. Note the important fact that since the structure of the porous matrix varies from plane to plane, this quenched disorder must be taken into account. 

%--------------------------------------------------------------------
\begin{figure}[ht]
\centering
\includegraphics[width=0.7\linewidth]{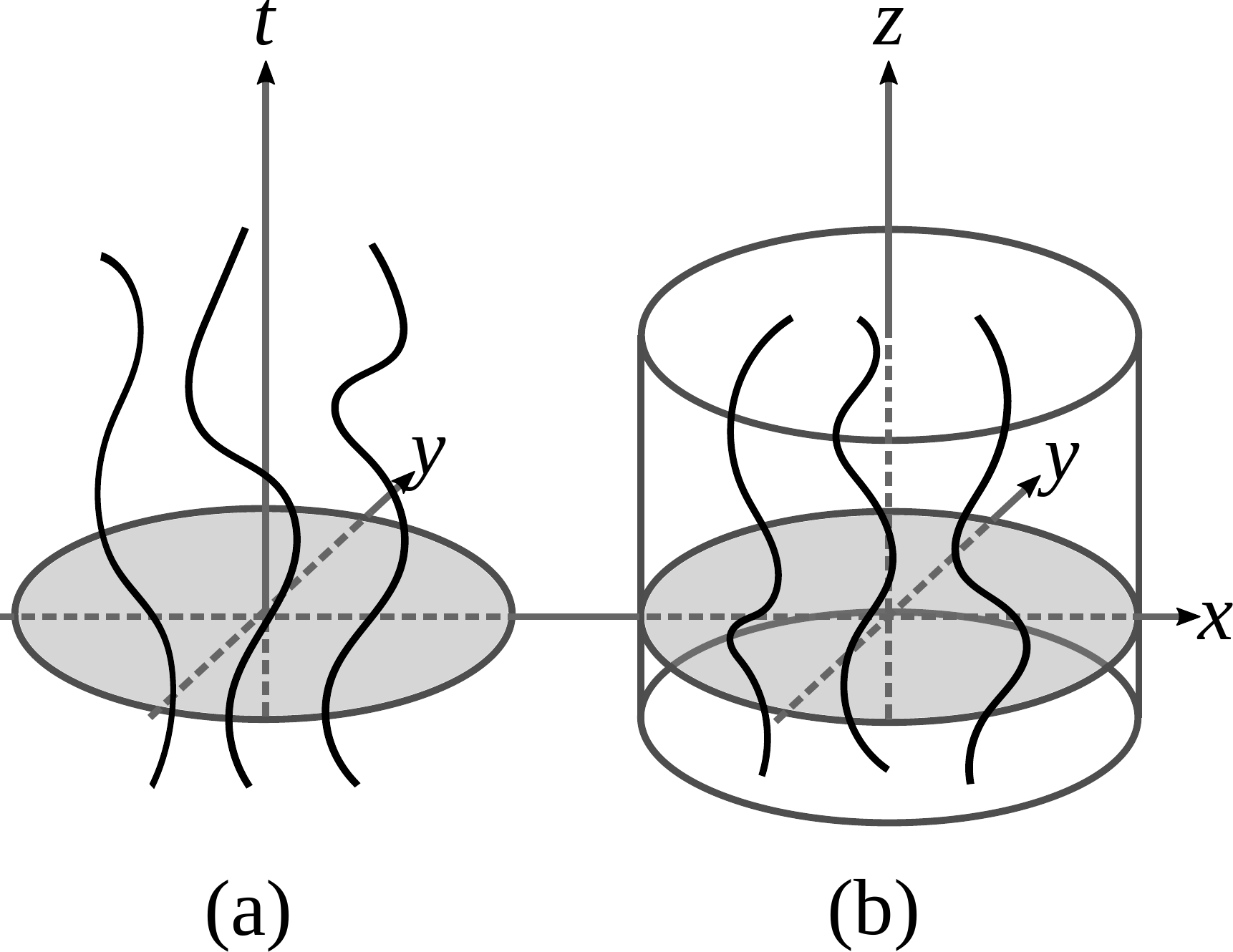}
\caption{We are illustrating to the left (a) the world lines of molecules in a two-dimensional gas in a space-time diagram. To the right (b), we show some streamlines of a fluid mixture flowing in a porous medium.}
\label{fig2-1}
\end{figure}
%--------------------------------------------------------------------

Before taking the next step, it is useful to think of the following system: We imagine a two-dimensional gas confined inside a box.  
The molecules of the gas move around incessantly. 
At a given moment, the position and velocity of each molecule will define an instantaneous gas configuration. 
The aim of statistical mechanics in this context is to provide the configurational probability density for these instantaneous configurations. There is no production of entropy in this system from the motion of the molecules.  It is in equilibrium.
However, we may represent the gas in a three-dimensional {\it space-time plot,\/} see Figure~\ref{fig2-1}(a).  Then, each molecule is represented by its world line and the configurations represent the world lines cutting through planes orthogonal to the time axis. 

Figure~\ref{fig2-1}(b) shows the streamlines of a fluid flowing through a porous medium. There is a striking analogy between these streamlines and the world lines of the molecules in the space-time plot of molecules of the two-dimensional gas,
when we interpret the $z$-axis in Figure~\ref{fig2-1}(b) as a ``time" axis. Figure~7 in Reference \cite{slmbm20} illustrates this point in more detail.
Cuts through the porous medium as shown in Figure~\ref{fig1} are then analogous to the snapshots of configurations of the gas molecules taken at different times.  The flow entropy defined in Equation~\eqref{eq-2} then corresponds to the entropy of the gas molecules, and as in the gas, there is no production of this flow entropy along the $z$-axis.  

The volumetric flow rate $Q$ has the same value for all planes orthogonal to the flow axis. Hence, with the flow axis acting as a ``time" axis, $Q$ is a conserved quantity along this axis. We may therefore interpret $Q$ as being analogous to the internal energy of the two-dimensional gas. 
Note that neither $Q_w$ nor $Q_n$ are conserved, only their sum $Q$ (Equation~\eqref{eq-1}) is.  However, both have well-defined averages.  The porous medium block of Figure~\ref{fig1} may be seen as an analog of a two-dimensional gas that does not exchange heat with its surroundings. In other words, it is the analog of a {\it microcanonical system.\/}

\begin{figure}[ht]
\centering
\includegraphics[width=0.7\linewidth]{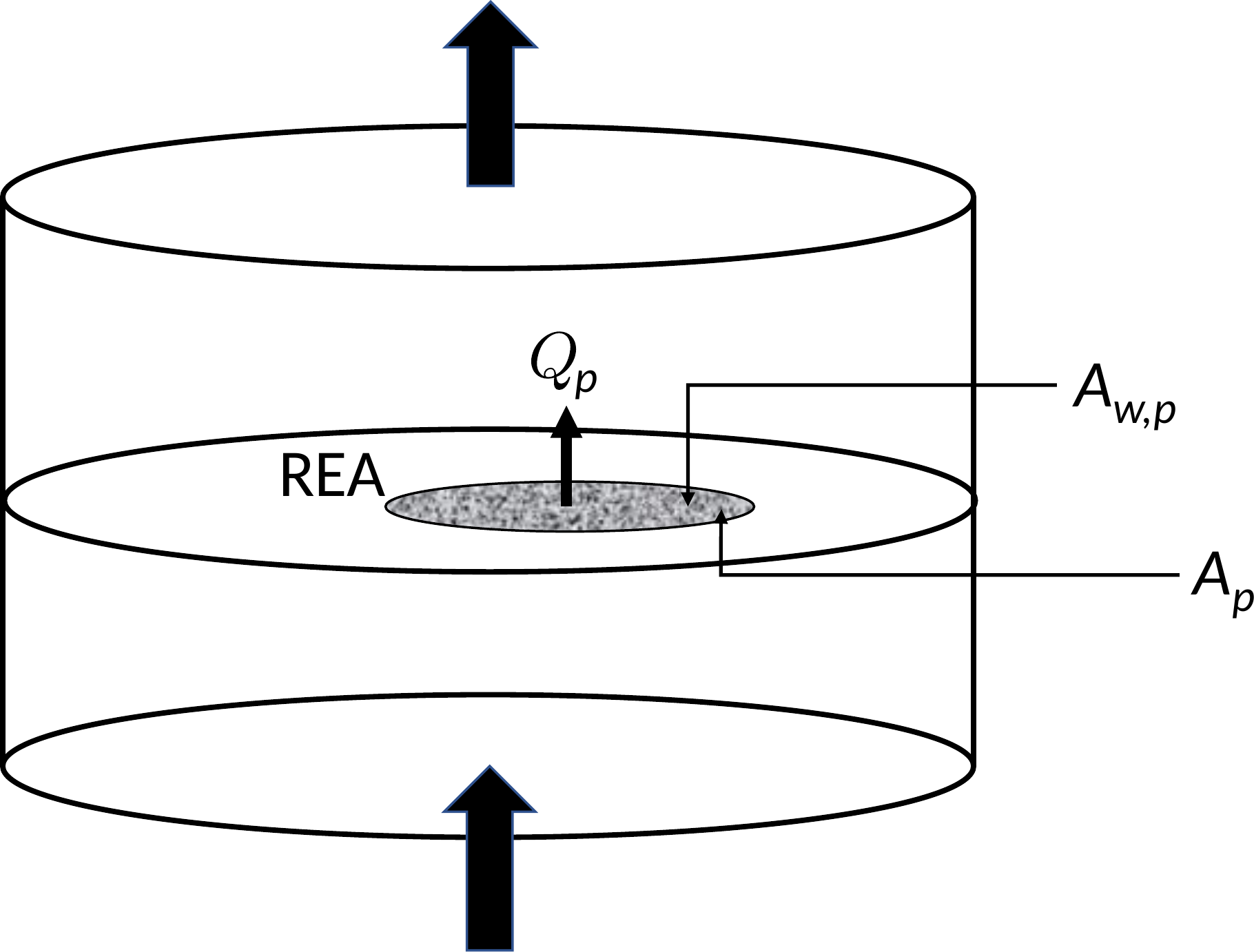}
\caption{Shows one of the planes cutting through the porous media block orthogonal to the flow direction which is upwards as marked with the arrows. In the plane, a sub-area that is large enough to reflect the behavior of the entire plane so that it acts as a Representative Elementary Area (REA) is selected. The REA is characterized by three variables besides its area $A$: The volumetric flow rate through it, $Q_p$, the area covered by the pores, $A_p$, and the area filled with the wetting fluid $A_{w,p}$.}
\label{fig2}
\end{figure}

Figure~\ref{fig2} shows one of the planes cutting through the porous medium orthogonally to the flow direction, i.e., the $z$-axis. 
A sub-area of this plane that is large enough to reflect the behavior of the entire plane is chosen. Hence, this area acts as an REA \cite{bb12}.  We characterize this REA by three variables in addition to its total area $A$: $Q_p$ the volumetric flow rate through it, $A_p$ which is the area inside the REA covered by pores, and $A_{w,p}$ which is the part of the pore area that is covered by the wetting fluid.  The configurations within the REA we refer to as  $X_p$. 
These configurations are a subset of the configurations $X$ in the entire plane.  Denoting $X_r$ as the part of $X$ which excludes the REA gives
\begin{equation}
\label{eq-3}
X=X_r\cup X_p\;.
\end{equation}

We now refer to the discussion in the Introduction (Section~\ref{intro}) and interpret the REA as the {\it system\/} and the plane excluding the REA as the {\it reservoir\/}. For the Jaynes maximum entropy approach to be applicable, we must have that
\begin{equation}
\label{eq-4}
P(X)=p_r(X_r)p(X_p)\;,
\end{equation}
where $p_r(X_r)$ is the configurational probability for the reservoir and $p(X_p)$ is the configurational probability for the REA.  The significance of this equation is that it ensures that it is possible to consider the REA as an autonomous system that interacts with the reservoir.  Without this property, a local description of the flow at the Darcy scale would then not be possible. 

%--------------------------------------------------------------------
%--------------------------------------------------------------------

It is the aim of this paper to verify the validity of Equation~\eqref{eq-4}.  This equation allows us to define a flow entropy for the REA, 
\begin{equation}
\label{eq-5}
\Sigma_p=-\int\ dX_p\ p(X_p)\ \ln p(X_p)\;,
\end{equation}
where the integral runs over all physically feasible configurations.

We maximize the entropy with the constraints that
the averages of $Q_p$, $A_p$ and $A_{w,p}$ are known.  This gives \cite{hfss22}
\begin{equation}
\label{eq-6}
p(X_p)=\frac{1}{Z}\ \exp\left[-\frac{Q_p(X_p)}{\theta}+\frac{\pi A_p(X_p)}{\theta}+ \frac{\mu A_{w,p}(X_p)}{\theta}\right]\;,
\end{equation}
where the partition function $Z$ is given by 
\begin{equation}
\label{eq-7}
Z(\theta,\pi,\mu)=\int dX_p\ e^{-Q_p(X_p)/\theta+\pi A_p(X_p)/\theta+\mu A_{w,p}(X_p)/\theta}\;.
\end{equation}
Here $Q_p(X_p)$, $A_p(X_p)$ and $A_{w,p}(X_p)$ are the variable values for the REA configuration $X_p$. 
It is through these three variables that contact is made with the pore-scale physics since this is where the
configuration $X_p$ enters. 
Three parameters appear in this equation: 1.\ the {\it agiture\/} $\theta$ which plays a role similar to that of temperature (and we note that the name, which is a contraction of the words ``agitation" and 
``temperature" has been
chosen to emphasize that this is {\it not\/} a temperature), 2.\ the {\it flow pressure\/} $\pi$ which is conjugate of the pore area $A_p$ --- and hence the porosity, and 3.\ the {\it flow derivative\/} $\mu$ which plays the role similar to the chemical potential and which is the conjugate to the wetting area $A_{w,p}$ and hence the wetting saturation 
$S_{w,p}=A_{w,p}/{A_p}$.

Equation~\eqref{eq-7} constitutes the scaling up from the microscopic level, in other words the pore level, to the Darcy level, since we may from it determine the values of the macroscopic variables.  We have thus succeeded in turning the scale-up problem from being a physical one to the mathematical problem of integration in Equation~\eqref{eq-7}.  The macroscopic variables that ensue from this approach are related through a thermodynamics-like formalism with all its richness \cite{hsbkgv18,hfss22}.  

In ordinary thermodynamics, one finds a set of {\it general\/} relations between the macroscopic variables.  They stem either from the Euler theorem for homogeneous functions or from the Gibbs relation \cite{c74,c91}.  The same applies to the present formulation of the two-phase flow problem.  We sketch the approach in the following.     

We define an average pore velocity $v_p=Q_p/A_p$ and an entropy 
density $\sigma_p=\Sigma_p/A_p$.  The average pore velocity $v_p$ depends on the flow entropy density $\sigma_p$ and the wetting saturation $S_{w,p}$: $v_p=v_p(\sigma_p,S_{w,p})$.  With these variables, we may construct an  equivalent to the {\it Gibbs relation,\/}
\begin{equation}
\label{eq-8-1}
dv_p=\theta\ d\sigma_p-\mu\ dS_{w,p}\;.
\end{equation}

We do a Legendre transform of the average flow velocity $v_p$ from $(\sigma_p,S_{w,p})$ to $(\sigma_p,\mu)$ as control variables, finding
\begin{equation}
\label{eq-8}
{\hat v}_n(\sigma_p,\mu)=v_p(\sigma_p,\mu)-S_{w,p}(\sigma_p,\mu)\mu\;,
\end{equation}
where we have defined the {\it thermodynamic non-wetting velocity\/} \cite{hsbkgv18}
\begin{equation}
\label{eq-9}
{\hat v}_n=\left(\frac{\partial Q_p}{\partial A_{n,p}}\right)_{A_{w,p},\sigma}\;.
\end{equation}
There is also the {\it thermodynamic wetting velocity\/}
\begin{equation}
\label{eq-10}
{\hat v}_w=\left(\frac{\partial Q_p}{\partial A_{w,p}}\right)_{A_{n,p},\sigma}\;.
\end{equation}
The non-wetting area $A_{n,p}$ is the area of the REA that is covered by the non-wetting fluid. We furthermore have that
\begin{equation}
\label{eq-11}
S_{w,p}=-\left(\frac{\partial {\hat v}_n}{\partial \mu}\right)_\sigma\;,
\end{equation}
and  
\begin{equation}
\label{eq-12}
\mu=-\left(\frac{\partial v_p}{\partial S_{w,p}}\right)_\sigma\;.
\end{equation}
Equations~\eqref{eq-8} through \eqref{eq-12} demonstrate the power of this approach.  These relations are far from obvious.   

There is one more central aspect that needs to be brought to light. The thermodynamic velocities defined in Equations~\eqref{eq-9}~and~\eqref{eq-10}
are {\it not\/} the pore velocities of the fluids
\begin{equation}
\label{eq-12-1}
v_w=\frac{Q_{w,p}}{A_{w,p}}\;,
\end{equation}
and  
\begin{equation}
\label{eq-12-2}
v_n=\frac{Q_{n,p}}{A_{n,p}}\;,
\end{equation}
where $Q_p=Q_{w,p}+Q_{n,p}$ in analogy with Equation~\eqref{eq-1}. 
Rather, they are related through the two equations \cite{hsbkgv18,rsh20,rpsh22}
\begin{eqnarray}
v_w={\hat v}_w-S_{w,p} v_m\;,\label{eq-13}\\
v_n={\hat v}_n+S_{n,p} v_m\;,\label{eq-14}
\end{eqnarray}
where $S_{n,p}=A_{n,p}/A_p=1-S_{w,p}$ and $v_m$ is the {\it co-moving velocity.\/}  It turns out from experimental and numerical data that the 
co-moving velocity is extraordinarily simple \cite{rpsh22},
\begin{equation}
\label{eq-15} 
v_m=a(\sigma)+b(\sigma)\mu\;,
\end{equation}
where $a(\sigma)$ and $b(\sigma)$ are functions of the flow entropy density.  There is no equivalent to the co-moving velocity in ordinary thermodynamics \cite{ph22}.

Calculating the partition function $Z(\theta,\pi,\mu)$ defined in Equation~\eqref{eq-7} requires a knowledge of the pore-scale configurations $X_p$ through the three variables $Q_p(X_p)$, $A_p(X_p)$ and $A_{w,p}(X_p)$.  Furthermore, the integral runs only over physically feasible configurations.  As already mentioned, this is where the characteristics of a given porous medium and the fluids enter. It is here details of the pore scale physics enters, such as interfacial tension gradients and interface curvature at the fluid-fluid interfaces.  This is where contact is made between this theory and the ongoing research on the pore-scale physics of immiscible two-phase flow.    

%--------------------------------------------------------------------
\section{Dynamic Pore Network Model}
\label{dpnm}

In order to explore the validity of Equation~\eqref{eq-4}, we use a 
dynamic pore network model \cite{sgvh19,jh12} originally developed by Aker et al.\ \cite{ambh98} and then further developed in e.g., \cite{kah02,rh06,toh12,gvkh18,gwh20,wgbkch20,fsrh21}, including direct comparison with experimental systems, \cite{sh17,zmpcv19}.  In the latter of these two references, the performance of the model is also compared to other models. 

\begin{figure}[ht]
\centering
\includegraphics[width=0.8\linewidth]{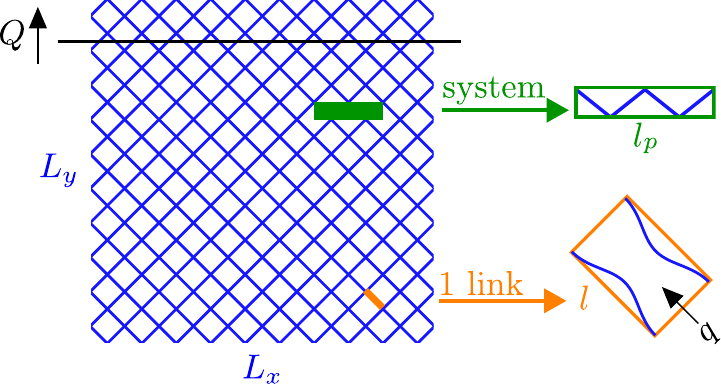}
\caption{Two dimensional dynamic pore network model with dimensions $L_x \times L_y$ links consists of hourglass shaped links with length $l$ and volumetric flow rate $q$ passing through them, oriented $45^{\circ}$ from the average flow direction.
The total volumetric flow rate $Q$ is constant over all the cross sections normal to the average flow direction.
An example of a ``system" with length $l_p=4$~links is marked, and the rest of the network surrounding the system is the ``reservoir." }
\label{fig3}
\end{figure}

We illustrate the model as it is implemented in the context of the present paper in Figure~\ref{fig3}. We use a square lattice where the links represent single pores, all having the same length $l$, but with a distribution in their radii. The lattice has dimensions $L_x\times L_y$ measured in units of $l$, and we implement periodic boundary conditions in both the flow direction and the transversal direction.  The square lattice is oriented at $45^\circ$ angle with respect to the average flow direction.

The links connecting neighboring nodes contain the pore throats. The nodes have no volume associated with them.  The variation in the cross-sectional area of the pore throat and pore bodies are modeled by an hourglass shape so that a fluid meniscus in link $i$ will generate a capillary pressure according to the Young-Laplace Equation~\cite{b17}
\begin{equation}
  \label{eq:pc}
  p_{c,i}(x) = \frac{2\gamma \cos{\theta}}{r_i(x)}\;,
\end{equation}
where $x\in [0,l]$ is the position of the interface along the center axis of the link, having a length $l$.  Here $\gamma$ is the surface tension and $\theta$ is the wetting angle measured through the wetting fluid which is the fluid that has the smallest angle with the solid wall.
We note that this expression is only valid under hydrostatic conditions. Hence, using it in a dynamic setting implies the assumption that the motion of the interfaces is slow.  This assumption is difficult to justify during Haines jumps. We still use it as an approximation that enters together with all the other approximations that the model requires. We furthermore ignore hysteretic effects associated with the wetting angle with the same justification as for using the Young-Laplace equation. The variable indicating the shape of the link in Equation~\eqref{eq:pc} is the radius of the link at position $x$, which is given by 
\begin{equation}
  \label{eq:radius}
  r_i(x) = \frac{r_{0,i}}{1-c\cdot \cos\left( \frac{2\pi x}{l} \right)}\;,
\end{equation}
where $c$ is the amplitude of the variation and $r_0/c$ is randomly chosen from the interval $[0.1 l, 0.4 l]$, thus creating a disorder in the properties of the network.

The fluids within a given link are pushed with a force caused by the total effective pressure across it which is the difference between the pressure drop between the two nodes it is attached to, $\Delta p$, and the total capillary pressure $\sum_k p_c(x_k)$ due to all the interfaces with positions $x_k \in [0,l]$. 
The model has been set to allow up to four interfaces in each link, and this necessitates merging of the interfaces as described in \cite{sgvh19}. 

The constitutive relation between the volumetric flow rate $q_i$ through link $i$ and pressure drop $\Delta p_i$ across the same link is \cite{sgvh19,w21}
\begin{equation}
  \label{eq:qi}
  q_i =
  - \frac{\pi \bar r_i ^4}{8\mu_i l} \left(\Delta p_i - \sum_k p_{c,i}(x_k)\right)\;,
\end{equation}
where ${\bar r}_i$ is the average hydraulic radius along the link. Furthermore, we have that
$\mu_i=s_{w,i}\mu_{w,i}+s_{n,i}\mu_{n,i}$ 
is the saturation-weighted viscosity of the fluids in link $i$ where $s_{w,i}=V_{w,i}/V_i$ and $s_{n,i}=V_{n,i}/V_i$ are the saturations of the wetting fluid and the non-wetting fluid respectively with viscosities $\mu_{w,i}$ and $\mu_{n,i}$, and  volumes $V_{w,i}$, $V_{n,i}$, and $V_i=\pi{\bar r}_i^2 l$ .

%--------------------------------------------------------------------
%--------------------------------------------------------------------

In order to calculate the flow through the links and move the interfaces correspondingly, we solve the Kirchhoff for the network using a conjugate gradient algorithm \cite{bh88}. Our numerical precision in determining the flow rates is $10^{-6}$. 

Using the terminology introduced in the Introduction, we divide the network into a ``system", corresponding to the REA, and a ``reservoir" which is the rest of the network.
The systems are chosen as illustrated with an example in Figure~\ref{fig3} where the system is placed orthogonally to the flow direction, i.e., in the same way as in Figure~\ref{fig2}. 
Systems are made up of $l_p$ number of links, for instance, the system in Figure~\ref{fig3} has $l_p=4$~links. 
The pore area $A_p$ of a system is the sum of the transverse area, the area orthogonal to the total flow direction, of each link belonging to that system,
\begin{equation}
\label{eq-20}
A_p=\sum_{i=1}^{l_p} \sqrt 2 \pi ({\bar r}_i)^2\;,
\end{equation}
where $\sqrt 2 = 1/\cos{(45^\circ)}$ comes from the fact that links in the model are oriented in $45^\circ$ angle from the flow direction.
Similarly, the total wetting fluid pore area of the system $A_{w,p}$ is the sum of the product of the transverse area and the wetting fluid saturation in each link,
\begin{equation}
\label{eq-20-2}
A_{w,p}=\sum_{i=1}^{l_p} \sqrt 2 \pi ({\bar r}_i)^2 s_{w,i} \;.
\end{equation}
The volumetric flow rate through a system with $l_p$ links is 
\begin{equation}
\label{eq-21}
  Q_p=\sum_{i=1}^{l_p} q_i\;,
\end{equation} 
and its wetting saturation is 
\begin{equation}
\label{eq-22}
S_{w,p}= \frac{ A_{w,p}}{A_p}\;.
\end{equation} 

%--------------------------------------------------------------------
\section{Numerical Investigations}
\label{system}

The simulations start from a random distribution of fluids within the network. 
The model is then integrated forwards in time while monitoring the pressure drop across it. 
When the pressure drop settles to a well-defined and stable average value, the model has reached steady-state flow. 
At this point, $20$ system locations are chosen randomly at every $100^\text{th}$ time-iteration, to get measurements that are mostly uncorrelated in time and space.
Within each of these systems, the values of $Q_p/l_p$ and $S_{w,p}$ are measured. 
This procedure ensures averaging not only over the motion of the fluids but also over the disorder of the porous medium itself. 
This process is repeated for a time corresponding to the passing of approximately $25$ pore volumes of fluid through the model, where pore volume is the total volume of the links in the model. 
We do this for different widths $l_p$ for the systems.
In addition, the changes in the model size are studied by testing various model widths $L_x$ while keeping the total length of the model fixed at $L_y = 60$~links. 
The links in the model are all $l=1$~mm long. 
The two immiscible fluids have $\gamma = 3.0\cdot 10^{-5}$~N$/$mm, $\mu_w=\mu_{nw}=0.1$~Pa$\cdot$s and $\theta=70^\circ$. 
The overall wetting saturation for the network is fixed at $S_w=0.5$.
Due to the periodic boundary conditions, the volume of the fluids is conserved and the total saturation is constant.
The total volumetric flow rate per unit width of the network is fixed at $Q/L_x=0.7$~mm$^3/($s$\cdot$link$)$.
The capillary number can be calculated from $\mathit{Ca}=\mu Q/ (\gamma A_\text{tot})$ where $A_\text{tot}$ is the total cross sectional area \cite{sgvh19}, giving $\mathit{Ca}\approx 0.012$. 
%% (0.1e-6)/(3e-5) * 0.7 / (pi*0.25^2) = 0.01188356908419485

%--------------------------------------------------------------------
\subsection{System Variable Statistics}
\label{sysVarStat}

\begin{figure}[ht]
\centering
\includegraphics[width=1.0\linewidth]{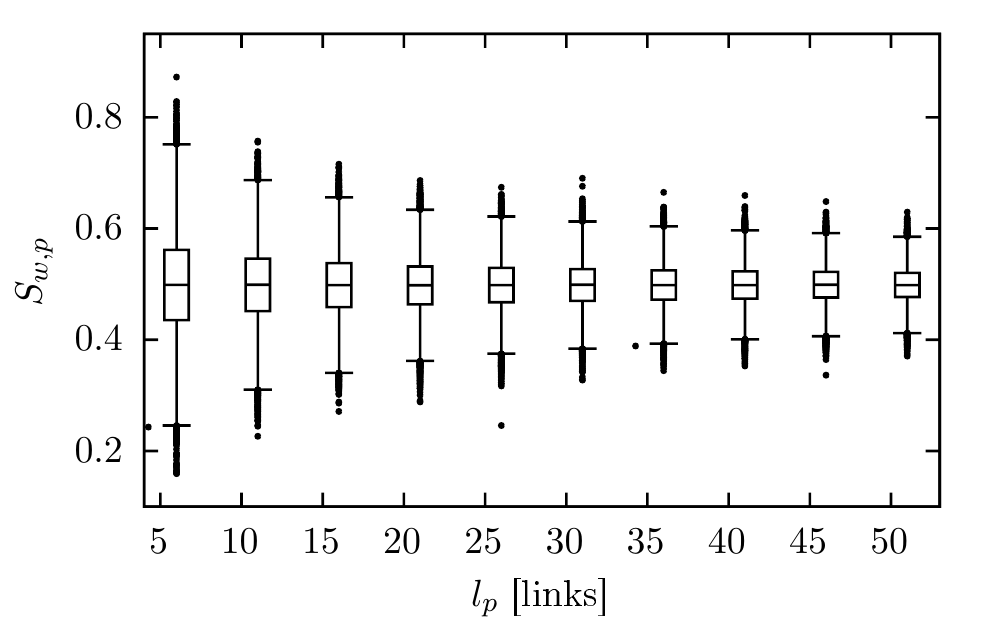}
\caption{Box plots showing the wetting fluid saturation $S_{w,p}$ in systems with width $l_p$. 
The model has dimensions $120\times 60$~links$^2$.}
\label{fig4}
\end{figure}

\begin{figure}[ht]
\centering
\includegraphics[width=1.0\linewidth]{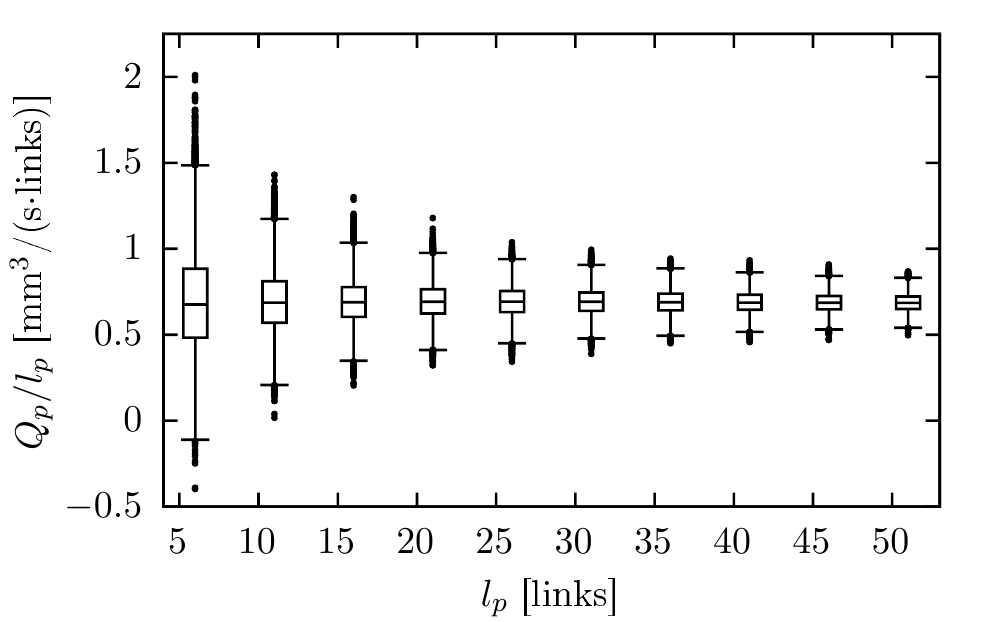}
\caption{Box plots showing the volumetric flow rate per unit system-width $Q_p/l_p$ in systems with width $l_p$.
The model has dimensions $120\times 60$~links$^2$.}
\label{fig5}
\end{figure}

Figures~\ref{fig4}~and~\ref{fig5} show box plots of $S_{w,p}$ and $Q_p/l_p$ as functions of $l_p$, for a network with dimensions $120\times 60$~links$^2$.
In each box plot, the lower edge of the box which we can denote $b_1$, the center line (median) $b_2$ and the upper edge $b_3$ correspond to the $25^\text{th}$, $50^\text{th}$ and $75^\text{th}$ percentiles of the data, respectively. 
The lower and upper limits that exclude the outliers are $b_1-1.5(b_3-b_1)$ and $b_3+1.5(b_3-b_1)$, respectively.

%--------------------------------------------------------------------
%--------------------------------------------------------------------

Since the control parameters for the entire network are fixed at $S_w=0.5$ and $Q/L_x=0.7$~mm$^3/($s$\cdot$link$)$, the average values $\langle S_{w,p}\rangle$ and $\langle Q_p/l_p\rangle$ in the systems will be the same as these values if measured with large enough statistics while the fluctuations around them will depend on $l_p$.
In Figures~\ref{fig4}~and~\ref{fig5}, the medians for all system sizes, $l_p$, agree well with these expected average values.
This factor indicates the existence of REAs since the intensive quantities inside REA must have well-defined averages that are independent of the size of the REA.
This agreement is even true for systems as small as $l_p=6$~links.

Furthermore, both Figures~\ref{fig4}~and~\ref{fig5} show a steady decrease in the variations in the distributions with increasing $l_p$ as the edges of the boxes approach the medians. 
This is more prominent for $Q_p/l_p$ in Figure~\ref{fig5} than for $S_{w,p}$ in Figure~\ref{fig4}. 
This is a reflection of $Q/L_x$ being constant for any cut through the model orthogonally to the average flow direction, since $Q_p/l_p \to Q/L_x$ as $l_p \to L_x$.
On the other hand, $S_{w,p}$ has no such restrictions and is therefore allowed to fluctuate, even when $l_p=L_x$.   
The fact that there is a smaller spread in both distributions with increased $l_p$ is another factor that signals possible REAs.

%--------------------------------------------------------------------
\subsection{Size Dependence of Fluctuations}
\label{stdev}

\begin{figure}[htpb]
  \centering
  \includegraphics[width=1.0\linewidth]{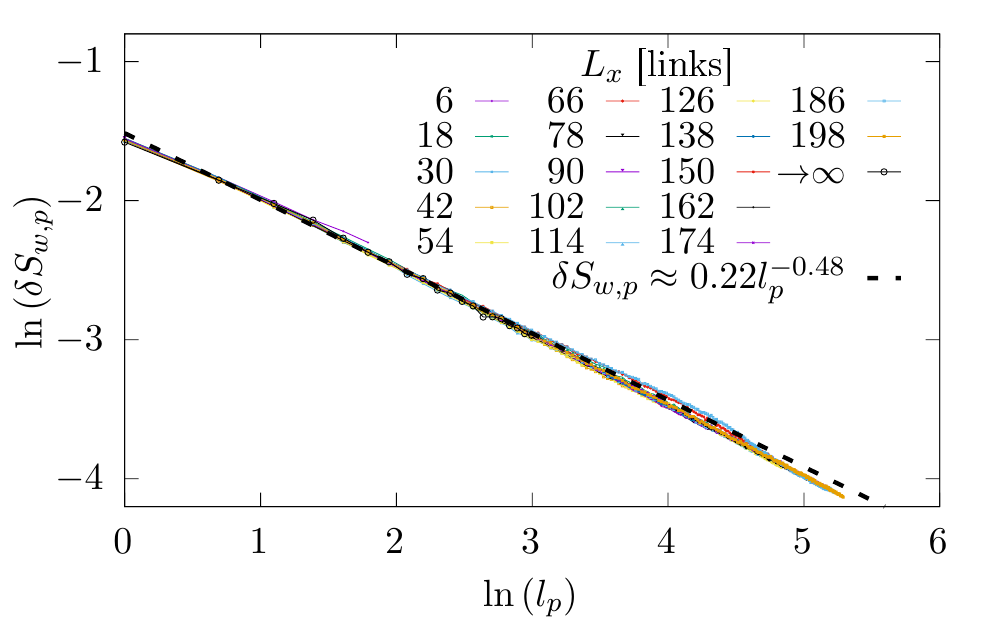}
  \caption{Standard deviation of the wetting fluid saturation $S_{w,p}$ in systems with widths $l_p$ residing in models having widths $L_x$ 
  and length $60$~links. }
  \label{fig6}
\end{figure}

\begin{figure}[htpb]
  \centering
  \includegraphics[width=1.0\linewidth]{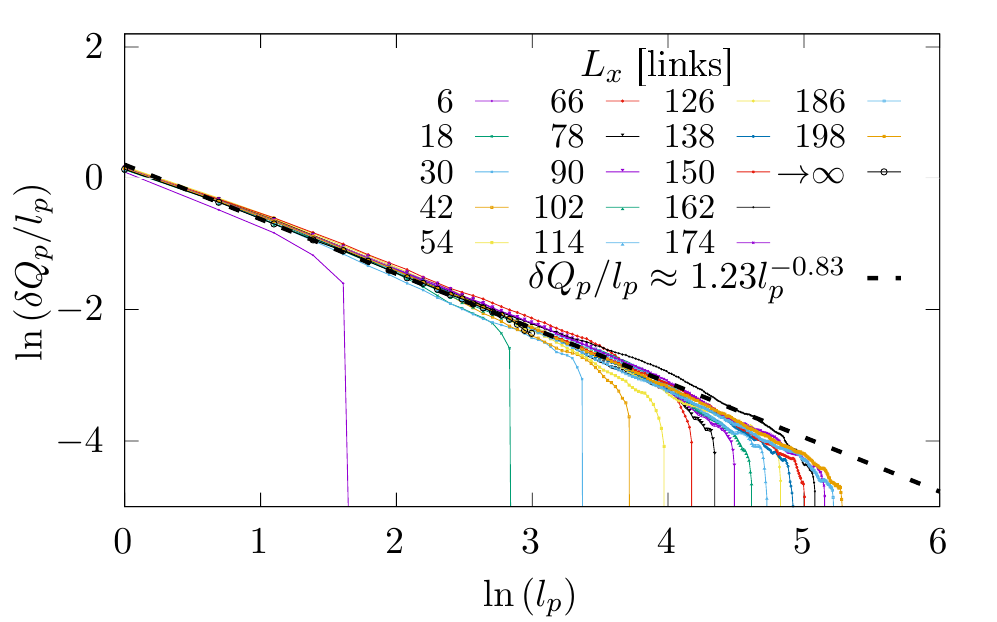}
  \caption{Standard deviation of the volumetric flow rate per unit system width $\delta Q_p/l_p$ in systems with widths $l_p$ residing in models with widths $L_x$ and length $60$~links. }
  \label{fig7}
\end{figure}

One way to quantitatively study the fluctuations in $\xi\in\left\{S_{w,p},Q_p/l_p\right\}$ is through their corrected standard deviations given by \cite{sas04}
\begin{equation}
  \label{eq:std}
  \delta \xi = \sqrt{ \frac{\sum_i^N \left(\xi_i-\langle \xi\rangle\right)^2}{N-1} }
\end{equation}
where $N$ is the number of measurements and $\langle \xi\rangle = \left(\sum_i^N \xi_i\right)/N$ is the mean. 
The standard deviations of $S_{w,p}$ and $\delta Q_p/l_p$ as a function of the system width $l_p$ are shown in Figures~\ref{fig6}~and~\ref{fig7}, respectively.


We note the difference in behavior in Figures~\ref{fig6}~and~\ref{fig7} in that $\delta Q_p/l_p$ drops off dramatically when $l_p$ approaches $L_x$ whereas no such effect is seen for $\delta S_{w,p}$.  
This is caused by the fact that $Q$ is not fluctuating in the planes orthogonal to the average flow direction, whereas there is no such constraint for $S_w$, which is a factor also mentioned earlier.
To avoid the measurements taken inside the systems being affected by the boundary effects, REA needs to be adequately smaller than the total model.

\begin{figure}[]
  \centering
    \includegraphics[width=1.0\linewidth]{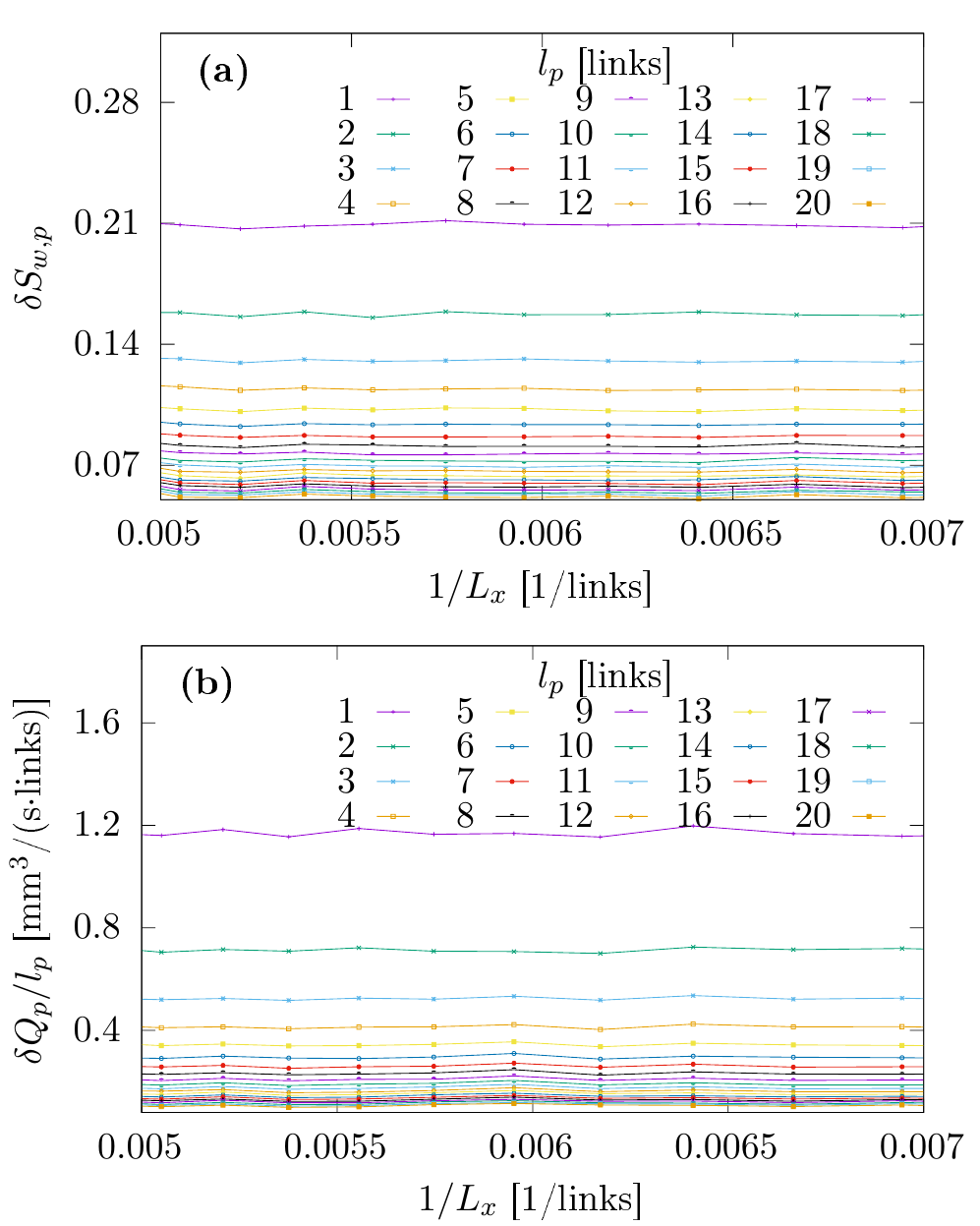}
  \caption{Standard deviation of (a) the wetting fluid saturation $S_{w,p}$ and (b) the volumetric flow rate per unit system width $Q_p/l_p$ in systems with widths $l_p$ residing in models having widths $L_x$ and length $60$~links.}
\label{fig8}
\end{figure}

We also show in Figures~\ref{fig6}~and~\ref{fig7} the results of extrapolation to infinitely large model $L_x\to\infty$. 
To understand how this was calculated, start by looking at Figure~\ref{fig8} where the results from Figures~\ref{fig6}~and~\ref{fig7} have been plotted in a different way.
In order to extrapolate to $L_x\to\infty$, the simulation results used must be from cases where the systems are much smaller than the model. 
To comply with this, the extrapolation process was performed for $l_p\in[1,20]$~links and $L_x\in[150,198]$~links. 
For these values of $l_p$ and $L_x$, Figure~\ref{fig8} shows that there is a linear relationship between $\delta\xi \in\left\{\delta S_{w,p},\delta Q_p/l_p\right\}$ and $1/L_x$.
Therefore, linear regression fit of the form 
\begin{equation}
  \label{eq:fitLxInftyType1prep}
  \delta \xi = \frac{c_1}{L_x}+ \delta \xi_\infty\;,
\end{equation}
can be performed for each $l_p$, where $c_1$ and $\delta\xi_\infty$ are constants.
It can be observed from Equation~\eqref{eq:fitLxInftyType1prep} that $\delta \xi\to \delta \xi_\infty$ when $L_x\to\infty$, hence $\delta \xi_\infty$ are the extrapolation results. 
In Figure~\ref{fig8}, $\delta\xi_\infty$ are the intersections the linear fits in Equation~\eqref{eq:fitLxInftyType1prep} make with the vertical axis.

After obtaining estimates for $\delta\xi_\infty$ for each $l_p$, we do a power law fit
\begin{equation}
  \label{eq:fitLxInftyType1}
  \delta\xi_\infty = c_2\ l_p^{-\beta}
\end{equation}
to model the relationship between $\delta\xi_\infty$ and $l_p$ where $c_2$ is a constant and $\beta$ is an exponent.
The result is
\begin{equation}
  \label{eq:fitLxInftyType1_SwpResult}
  \lim_{L_x\to\infty} \delta S_{w,p} \approx  0.22 l_p^{-0.48}
\end{equation}
for Figure~\ref{fig6} and 
\begin{equation}
  \label{eq:fitLxInftyType1_QpApResult}
  \lim_{L_x\to\infty} \left(\frac{\delta Q_p}{l_p}\right) \approx  1.23 l_p^{-0.83}
\end{equation}
for Figure~\ref{fig7}.

Based on the central limit theorem, the standard deviation of the average of $l_p$ equally distributed independent variables is proportional to ${l_p}^{-1/2}$. 
The quantities $S_{w,p}$ and $Q_p/l_p$ are both intensive quantities representing averages in $l_p$.
We note from Equation~\eqref{eq:fitLxInftyType1_SwpResult} that this is the case with $S_{w,p}$, which could indicate that samples are uncorrelated.
However, that the fluctuations $\delta Q_p/l_p$ in Equation \eqref{eq:fitLxInftyType1_QpApResult} scales as one over $l_p$ to the power 0.83 is a surprise presumably indicative of the samples being non-zero correlated in such a way that they fall off faster than when there are no correlations. 
This further means that the reservoir should be larger than the spatial correlation length for the systems to be not affected by finite-size effects.

%--------------------------------------------------------------------
\subsection{Reservoir Independence}
\label{independence}

We have now reached the central aim of this paper: Testing the validity of Equation~\eqref{eq-4} for our dynamic pore network model.  
This is done by keeping $l_p$ fixed and varying $L_x$ while monitoring histograms of $S_{w,p}$ and $Q_p/l_p$. 
If Equation~\eqref{eq-4} is valid for this model, the histograms should be independent of the model size $L_x$ for large enough $L_x$.  

\begin{figure}[]
  \centering
  \includegraphics[width=1.0\linewidth]{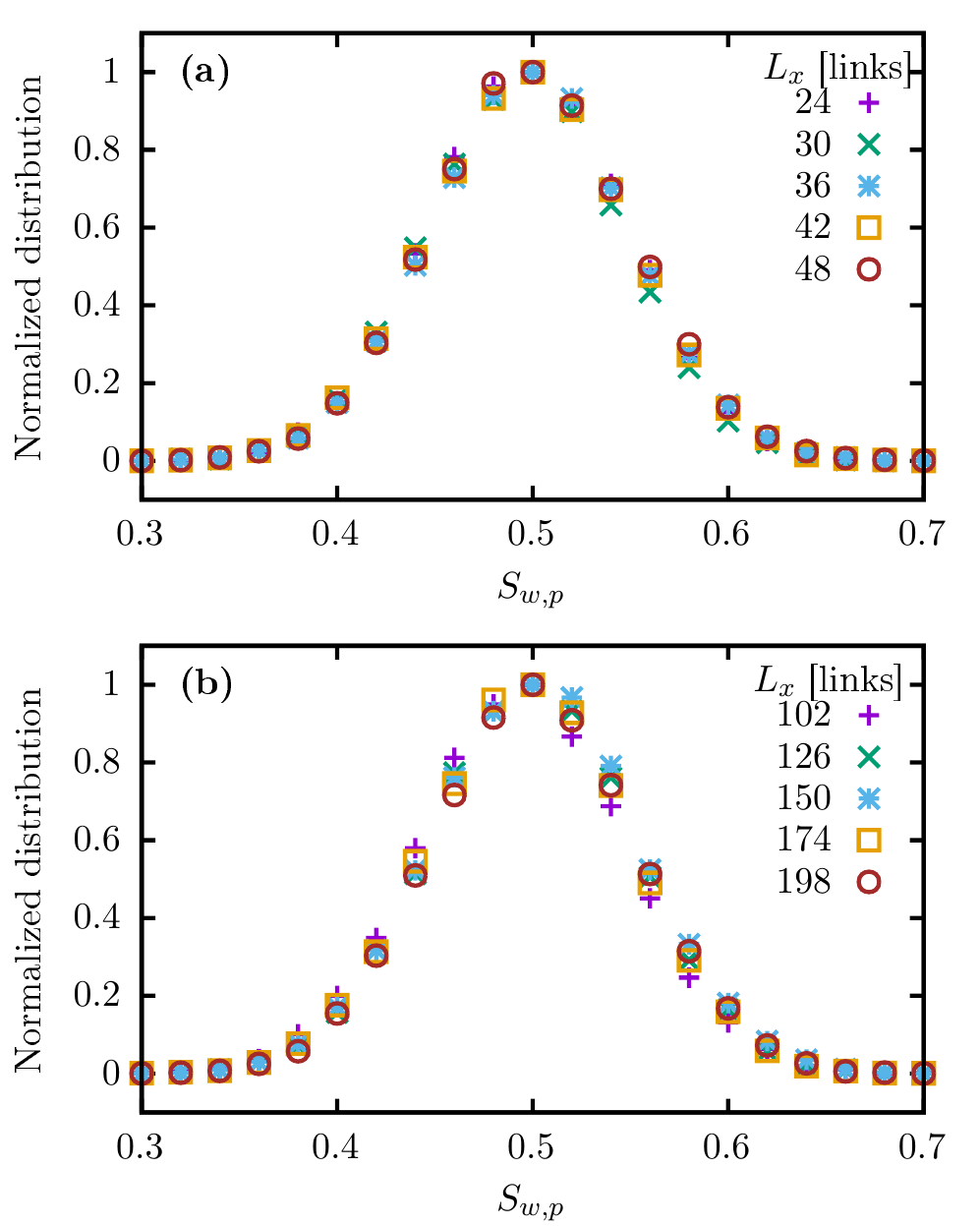}
  \caption{Normalized histogram for the wetting fluid saturation $S_{w,p}$ in systems with width $l_p=20$~links.
   The model has length $60$~links and widths $L_x$ that are close to $l_p$ in (a) and are much larger than $l_p$ in (b).}
  \label{fig9}
\end{figure}

\begin{figure}[]
  \centering
  \includegraphics[width=1.0\linewidth]{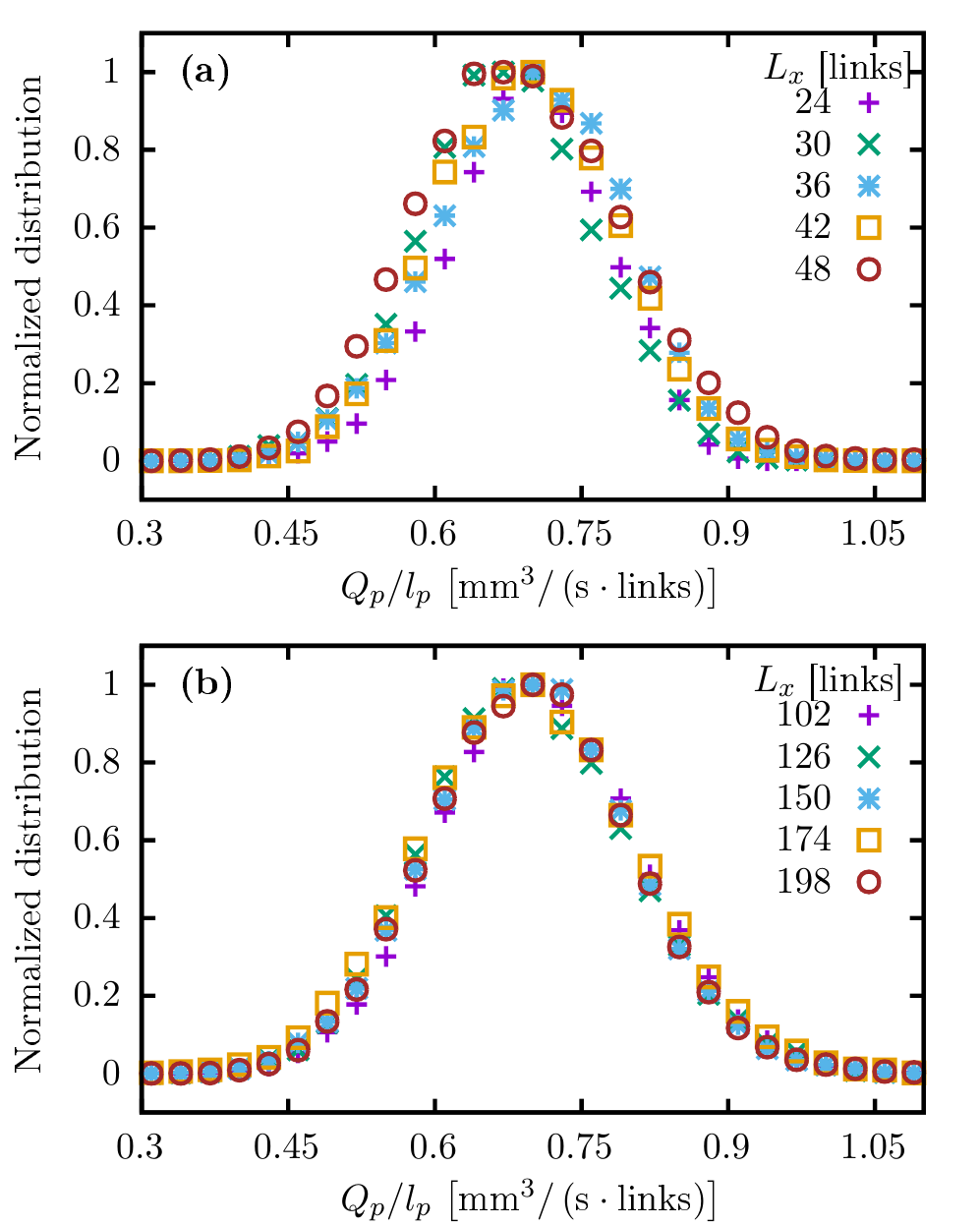}
  \caption{Normalized histogram for the volumetric flow rate per link $Q_p/l_p$ in systems width $l_p=20$~links.
   The model has length $60$ links and widths $L_x$ that are close to $l_p$ in (a) and are much larger than $l_p$ in (b).}
  \label{fig10}
\end{figure}

The normalized histograms of $S_{w,p}$ and $Q_p/l_p$, measured for systems of width $l_p =20$ links, are shown in Figures~\ref{fig9}~and~\ref{fig10} respectively. 
We have split the two figures into two sub figures each in order to increase readability. 
Figures~\ref{fig9}(a)~and~\ref{fig10}(a) show the normalized histograms for $L_x$ being close to $l_p$, whereas Figures~\ref{fig9}(b)~and~\ref{fig10}(b) show the normalized histograms for $L_x$ much larger than $l_p$. 

The normalized histograms for $S_{w,p}$ in Figure~\ref{fig9} seem to overlap for essentially all values of $L_x$.
This effect can also be seen in the standard deviations in Figure~\ref{fig6} where $\delta S_{w,p}$ approximately follows Equation~\eqref{eq:fitLxInftyType1_SwpResult} regardless of the model size $L_x$ or the system size $l_p$.
This means, in the case of $S_{w,p}$, the reservoir independence seems to be satisfied regardless of the difference between $l_p$ and $L_x$.

On the other hand, the histograms of $Q_p/l_p$ differ more from each other when $L_x$ is close to $l_p$, see Figure~\ref{fig10}(a), compared to $L_x$ being much larger than $l_p$ where they overlap significantly larger, see Figure~\ref{fig10}(b).
This behavior is reflected in the standard deviation results in Figure~\ref{fig7} as well where $\delta Q_p/l_p$ following Equation~\eqref{eq:fitLxInftyType1_QpApResult} only when $l_p$ is less than $L_x$.  
From these findings, we conclude that $Q_p/l_p$ is independent of the reservoir size when $L_x$ is sufficiently larger than $l_p$. 
This difference in behavior is presumably related to the flow rate $Q/L_x$ being constant in all layers whereas the wetting saturation $S_w$ fluctuates.  


The results combined indicate that reservoir independence is valid for our model when the reservoir is adequately larger than the system.

%--------------------------------------------------------------------
\section{Conclusion}
\label{conclusion}

The aim of this paper has been to address the plausibility of a necessary condition for the Jaynes statistical mechanics formulation \cite{j57} to be applicable to immiscible and incompressible two-phase flow in porous media. 
The condition demands that a such porous medium can be split into a system, functioning as a Representative Elementary Area (REA), and a reservoir as in ordinary thermodynamics. This requires the statistics of the system to be independent of the size of the reservoir. Using dynamic pore network model simulations, we studied this by measuring distributions of key parameters using systems and reservoirs with different sizes. 

First, the results show that there exist systems that can qualify as REAs within which the studied distributions have small spread and have well defined averages independent of the size of the REAs. 

Second, REAs exhibit reservoir independence as demonstrated in Figures~\ref{fig9}~and~\ref{fig10}. Hence, the 
central Equation~\eqref{eq-4} 
\begin{equation}
\nonumber
P(X)=p_r(X_r)p(X_p)\;,
\end{equation}
works for the dynamic pore network model.    
   
As was alluded to at the end of the Introduction, the importance of the validity of Equation~\eqref{eq-4} goes beyond verifying the Jaynes statistical mechanics framework \cite{hfss22}.  If Equation~\eqref{eq-4} would have failed, any attempt at constructing a local theory for immiscible two-phase flow at the Darcy scale would be in jeopardy.  By local we mean that we can define variables that depend on a given point in the porous medium and the theory then provides relations between these variables depending only on that point. Relative permeability theory is an example of such a local theory.  A failure of Equation~\eqref{eq-4} would presumably necessitate the relations between variables containing integration over space.        

When the reservoir size approaches to infinity, as the extrapolation results show, the fluctuations in the wetting fluid saturation depend on the system size through an exponent of $-0.48$.
The same exponent in the case of volumetric flow rate per unit system width is $-0.83$.
The fact that at least one of the fluctuations corresponds to an exponent significantly different from $-1/2$ indicates a non-zero spatial correlation between the links in the network, according to the central limit theorem.
We speculate that this may be a consequence of total volumetric flow rate in the planes orthogonal to the average flow direction being a conserved quantity, whereas saturation in the same planes is not. 

The measured distributions of saturation and volumetric flow rate inside systems are more similar when reservoir is much larger than system than when system and reservoir are closer in size. 
The cases with similar distributions indicate that reservoirs in these cases are adequately larger than the spatial correlation length for the systems to be unaffected by finite-size effects. 
Reservoir independence can be said to be achieved in these cases. 

Our dynamic pore network model is capable of modeling porous media with a large number of pores (links). 
This comes at the cost of a simplified description of the structure of the pores and the motion of the fluids. 
Other models such as the Lattice Boltzmann Model \cite{rbt19} are capable of modeling the structure of the pores and the motion of the fluids inside them quite accurately. 
However, the price for this is the number of pores that may be modeled is limited. 
Nevertheless, attempts should be made to test reservoir independence of the systems also within the limits of this model. Another formidable task would be to analytically derive reservoir independence using hydrodynamics at the pore level.  

%--------------------------------------------------------------------

\section{Declarations}%
\label{sec:declaration}

\noindent \textbf{Author Contributions}:
All the authors contributed in developing the theory and the methodology and writing the manuscript to its final form.
HF performed the numerical simulations and data analysis . 
HF and SS wrote the code for the model. 

\noindent \textbf{Funding}: 
This work was supported by the Research Council of Norway through its Center of Excellence funding scheme, project number 262644.

\noindent \textbf{Acknowledgment}: 
The authors thank Carl Fredrik Berg, Eirik G.\ Flekk{\o}y, Daan Frenkel, Federico Lanza, H{\aa}kon Pedersen and Per Arne Slotte for interesting discussions. 

%%============================================================================%%
\bibliographystyle{abbrvnat}

%--------------------------------------------------------------------

%--------------------------------------------------------------------
\end{document}